\documentclass[preprint]{aastex6}

\usepackage[normalem]{ulem}

\renewcommand{\bm}[1]{\mbox{\boldmath{$#1$}}}

\def\rB{{\mathrm{B}}}
\def\rF{{\mathrm{F}}}

\def\a{{\bm{a}}}
\def\b{{\bm{b}}}
\def\v{{\bm{v}}}
\def\c{{\bm{c}}}
\def\h{{\bm{h}}}
\def\x{{\bm{x}}}

\def\hatbB{{\hat{\bm{B}}}}
\def\hatB{{\hat{B}}}

\def\be{\begin{eqnarray}}
\def\ee{\end{eqnarray}}

\usepackage{graphicx}

\usepackage{xspace}
\def\qsl{{\fontfamily{pcr}\selectfont
QSL} {\fontfamily{pcr}\selectfont Squasher}\xspace}

\shorttitle{QSL Squasher}
\shortauthors{Tassev \& Savcheva}

\begin{document}

\title{\qsl: A Fast Quasi-Separatrix Layer Map Calculator}

\author{Svetlin Tassev\altaffilmark{1,2}, Antonia Savcheva\altaffilmark{1}}
\affil{\altaffilmark{1}Harvard-Smithsonian Center for Astrophysics, 60 Garden Street,
Cambridge, MA 02138, USA \\ 
\altaffilmark{2}Braintree High School, 128 Town Street, Braintree, MA 02184, USA}
\email{svetlin.tassev@cfa.harvard.edu}
\begin{abstract}

Quasi-Separatrix Layers (QSLs) are a useful proxy for the locations where current sheets can develop in the solar corona, and give valuable information about the connectivity in complicated magnetic field configurations. However, calculating QSL maps even for 2-dimensional slices through 3-dimensional models of coronal magnetic fields is a non-trivial task as it usually involves tracing out millions of magnetic field lines with immense precision. Thus, extending QSL calculations to three dimensions has rarely been done until now. In order to address this challenge, we present \qsl  \ -- a public, open-source code, which is optimized for calculating QSL maps in both two and three dimensions on GPUs. The code achieves large processing speeds for three reasons, each of which results in an order-of-magnitude speed-up. 1) The code is parallelized using OpenCL. 2) The precision requirements for the QSL calculation are drastically reduced by using perturbation theory. 3) A new boundary detection criterion between quasi-connectivity domains is used, which quickly identifies possible QSL locations which need to be finely sampled by the code. That boundary detection criterion relies on finding the locations of abrupt field-line length changes, which we do by introducing a new Field-line Length Edge (FLEDGE) map. We find FLEDGE maps useful on their own as a quick-and-dirty substitute for QSL maps. \qsl allows constructing high-resolution 3D FLEDGE maps in a matter of minutes, which is two orders of magnitude faster than calculating the corresponding 3D QSL maps. We include a sample of calculations done using \qsl to demonstrate its capabilities as a QSL calculator, as well as to compare QSL and FLEDGE maps.

\end{abstract}

 
\section{Introduction} \label{sec:Intro}

Many important questions in solar physics concern phenomena that take place in the low-$\beta$ environment of the corona, such as flares and coronal mass ejections (CMEs), active region (AR) evolution and dynamics, heating of the corona and sources of the solar wind. For studying these, it is often useful to have a model of the 3D magnetic field in the corona as it still cannot be observed  and mapped directly. Such models can be potential \citep{Jiang12a}, linear force-free field \citep[LFFF; constant-$\alpha$; e.g.][]{Demoulin94, Abramenko96, Jiang12a}, or non-linear force-free field \citep[NLFFF; $\alpha(r)$; e.g.][]{vanBallegooijen04, Wiegelmann04, Valori05, Wheatland06, Schrijver06, Malanushenko12, Inoue12, Jiang12b}. Potential field source surface models have been in use for a long time and although well representative of the structure of the field at large heights in the corona, they by definition lack currents, and hence free energy, which is important for powering solar eruptions. An alternative are NLFFFs, which have gained significant popularity in recent years with the advent of numerous methods for their computation, which use either line-of-sight or vector photospheric magnetograms to produce a model of coronal magnetic fields or to extrapolate the observed photospheric magnetic field to the corona. However, these 3D magnetic fields are intrinsically complicated and although basic topological features, such as null points (NPs), fan-spine surfaces and flux ropes (FRs) can often be approximately identified just by inspecting field line plots, often, there is a need of quantitative topological analysis in order to make sense of the complicated 3D coronal magnetic field structure, its dynamics, and evolution.

Topological features in 2D and 2.5D, such as NPs \citep[e.g.][]{GorbatchevSomov88, Parnell10}, separatrices \citep{GorbatchevSomov88}, separator field lines, and null lines have been explored in solar physics context since the 80s. They are known to separate the field in connectivity domains. However, in the mid-90s a new topological term arose, namely quasi-separatrix layers \citep[QSLs;][]{Priest95, Demoulin96b}, which are the 3D generalizations of the above-mentioned features, now separating the field into quasi-connectivity domains. While the linkage of magnetic field lines over separatrices and NPs is discontinuous, across QSLs it is continuous but drastically changes. 

In the early description of \cite{Demoulin96a}, the strength of QSLs, i.e. the amount of the change in field line linkage, is measured by the norm of the Jacobian of the mapping of neighboring field lines \citep{1994ApJ...437..851L} from one end of the photosphere to the other. However, this quantity is not invariant with respect to the direction of tracing of the field lines. Consequently, \cite{Titov02} came up with an alternative covariant quantity quantifying QSL strengths, called the squashing factor, $Q$.

There is no general relationship between QSLs and electric currents. For example one can imagine a potential field with no currents and very strong QSLs (large $Q$ value), even separatrices. However, when applying boundary motions, strong electric current develop preferentially at QSLs whatever the footpoint motions are \citep{Aulanier05b}. Then, QSLs (as well as NPs and separatrices) are preferential sites for build-up of current sheets in the presence of footpoint motions, and hence are preferential sites where reconnection can take place. In addition, starting from a thick QSL distribution the evolution of the configuration can thin the QSLs, and hence the current sheets, down to the dissipative scales. Then, \cite{Aulanier05b} proposed that for non-eruptive flares the broader the initial QSLs, the larger the magnetic energy which can be stored before being released by reconnection. QSLs and related current sheets also have a key role in eruptive flares \citep{Aulanier10} and in coronal heating \citep{DemoulinP97}. This makes these topological features very important for studies of storage and release of magnetic free energy in the process of reconnection at all scales.

Quantitative studies of topology by deriving QSL maps in 2D from potential, LFFFs, and NLFFFs have been used over the past decade to tackle many problems in solar physics. The existence of a QSL wrapping around the flux rope and crossing itself at a very high-$Q$ topological feature, a hyperbolic flux tube \citep[HFT;][]{Titov07, Savcheva12a, Savcheva12b, Zhao14, Liu14}, has become the basic feature in the standard flare model in 3D \citep{Aulanier12, Janvier13}, confirmed by observations \citep{Janvier14, Savcheva15, Savcheva16a, Janvier16, Zhao16}. In this picture, tether-cutting reconnection happens at the HFT under the FR between J-shaped oppositely directed field lines, which slip \citep{Aulanier06b} over the photospheric traces of the HFT \citep{Janvier13} and produce S-shape field lines that feed the FR and post-flare arcade. This scenario was put forward supported by data-constrained NLFFF models and MHD simulations by \cite{Savcheva12b}. In this picture, the photospheric traces of the HFT are 2J-shaped \citep{Titov07, Aulanier10} and they match the 2J-shaped flare ribbons of classical two-ribbon flares \citep{Chandra09,Schrijver11}. The match between the shapes of QSLs and flare ribbons has been achieved recently by \cite{Liu14}, \cite{Savcheva15}, and \cite{Zhao16}. These QSLs have been shown to move together with the flare ribbons in direction perpendicular to the polarity inversion line (PIL) \citep{Savcheva16a, Janvier16}. 

The QSLs derived in \cite{Savcheva16a} and \cite{Janvier16} have been derived based on NLFFFs constrained only by pre-flare observations (magnetograms, and EUV and X-ray images), but have managed to reproduce the flaring topology and its evolution to a large extent. That indicates that these kinds of studies have potential predictive power as the use of topology analysis can show us the likely sites of flare reconnection a few hours before the event, as shown in \cite{Savcheva12a}. One could imagine going further and using the flare ribbon information and QSLs to work backwards and improve the initial NLFFF, thus providing better initial conditions for global data-driven MHD simulations of CME initiation and propagation \citep{Savcheva17}.

Further studies  show the evolution of QSL maps of solar ARs over several days. Some note the effects of flux cancellation on building sigmoidal flux ropes -- for example, see \cite{Savcheva12a}, who showed the transition from bald-patch separatrix surfaces \citep[BPSS;][]{Titov93} to a HFT. Others focus on the effects of quadrupolar topology on the possible breakout scenario \citep{Zhao14} or the effects of flux emergence on the development of a fan-spine NP topology \citep{Jiang16}. The global topology of active regions before eruption has been shown to be important for the characteristics of the dynamics, be it an eruption or just loop  reconfiguration \citep{Janvier16, Jiang16, Pontin16, Chintzoglou16}. 

Knowing the locations, extent, shape, and connections between connectivity domains, and the features they contain or border, can prove vital for understanding links between seemingly unconnected faraway regions on the Sun that erupt sequentially or almost simultaneously, i.e. sympathetic eruptions. A detailed study of one such event (1-2 August, 2010) was conducted by \cite{Titov12}, who showed that filaments embedded in neighboring pseudostreamers are activated sequentially after the first filament erupts and destabilizes the system \citep{Torok11}. Even if it is a single CME, the potential of the CME to have a large longitudinal extent or to present with a significant energetic particle signature at any point in the heliosphere is most probably dependent on the specifics of the global 3D topology in the corona and heliosphere as the CME evolves and propagates \citep{Masson13}. As a related phenomenon, the propagation of EUV dimmings may also turn out to be dependent on the global solar topology, neighboring the directly related AR \citep{Downs16}.

On a smaller scale, reconnection at QSLs have been potentially found important for the heating of the solar corona \citep{DemoulinP97,Schrijver10}. Reconnection in loop braiding has been theoretically and numerically explored for this purpose as well \citep{WilmotSmith09a, Pontin15}. QSLs in the outskirts of ARs  have been shown to drive plasma outflows \citep{Baker09} as evidenced by blueshifts in Hinode/EIS velocity maps of ARs, which could be important for understanding the outflow of plasma from the corona that contributes to the slow solar wind. Potential solar wind sources can be further derived by means of the S-web model of \cite{Antiochos11,Titov11,Linker11}, which utilizes QSLs at the source surface and below to look at the connectivity domains surrounding active regions and coronal holes, as well as the connections between them.   

Ultimately, with the speed-up and automation of NLFFF codes and QSL computation methods, we will be able to implement 3D QSL analysis in space weather predictive operations aimed at identifying the next likely region to erupt, studying the effect of the propagation of the CME ejecta and its particles, and predicting the direction and sign of the CME magnetic field when it reaches the Earth's magnetosphere. One step on this path is obtaining a fast, reliable 3D QSL code that can work on the whole Sun or in an AR in great detail. Such codes have been developed and used before for analyzing potential coronal magnetic field models \citep[][]{0004-637X-806-2-171}, as well as experimental flux rope configuration \citep[][]{PhysRevLett.103.105002}, yet they were never made public.

In this paper, we introduce a fast, freely-available, open-source code, \qsl, aimed at calculating 3D QSL maps, whose development was motivated by several potential uses, such as:

\begin{itemize}
\item Studying large resolution QSL physics and its application to reconnection theory.
\item Exploring large parameter spaces of possible topologies.
\item 3D studies of active region evolution, CME initiation and propagation.
\item Obtaining the evolution of topology over large periods of time with high cadence from data-driven or idealized MHD simulations at a wide range of scales.
\end{itemize}

The paper is organized as follows. In Section~\ref{over} we give an overview of the code. In Section~\ref{algo} we give details about the algorithm used in \qsl. We show illustrative results in Section~\ref{ex} and give our concluding remarks in Section~\ref{summary}.

\section{Code Overview}\label{over}

\qsl (Zenodo \doi{10.5281/zenodo.207471}) is written in C++ and depends on the Boost\footnote{\url{http://www.boost.org/}} and VexCL\footnote{\url{https://github.com/ddemidov/vexcl}} \citep{vexcl} libraries, on a working OpenCL\footnote{\url{https://www.khronos.org/opencl/}} implementation, as well as on their respective dependencies. The visualization scripts require Python with SciPy \citep{scipy} and PyEVTK\footnote{\url{https://www.python.org/}, \url{https://www.scipy.org/}, \url{https://bitbucket.org/pauloh/pyevtk}}. The code is intended to be run on a graphics processing unit (GPU). However, it can be run multithreaded on a CPU if one uses the POCL\footnote{\url{http://portablecl.org/}} \citep{pocl} OpenCL implementation.

The input for \qsl is 3D cubes containing the values of the magnetic field components sampled on a rectilinear grid in either Cartesian or spherical coordinates. For the exact file structure, we refer the reader to the manual distributed with the code.

The output of the code can be 2D or 3D arrays of $Q$ values, depending on whether the code is run in 2D or 3D mode to produce slices or data cubes, respectively. For slices, the output can be rendered as an image using the provided Python script. The output of 3-dimensional calculations is exported to VTK format, which can then be visualized using Paraview, VisIt or Mayavi among many\footnote{\url{http://www.vtk.org/}, \url{http://www.paraview.org/}, \url{https://wci.llnl.gov/simulation/computer-codes/visit/}, \url{http://docs.enthought.com/mayavi/mayavi/}}.

When the desired output is a slice through the volume of interest, the slices can have two types of geometry: planar or spherical. The code supports constructing planar slices of arbitrary orientation. In this case, one needs to specify the orientation, center, and axes span of the slice. Spherical slices are slices at a specified fixed radius, spanning a given range in latitude and longitude. 

\section{Algorithm}\label{algo}

Below we write down the equations solved by \qsl. The code can work in both Cartesian and spherical coordinates. Whenever we find it useful, we quote the explicit equations solved by the code for spherical coordinates.

\subsection{Integrating field lines}

The magnetic field lines $\x(\lambda)$ (where $\lambda$ is an affine parameter) are calculated as the integral curves of the unit magnetic field, $\hatbB$.  Thus, in Cartesian coordinates, we have:
\be\label{fl}
\partial_{\lambda}\x(\lambda)=\hatbB(\x(\lambda))\ .
\ee
In spherical coordinates, the field lines are given as solutions to (keeping the $\lambda$ dependence explicit):
\be\label{fl_sph}
r(\lambda)\cos\Big(\theta(\lambda)\Big)\partial_\lambda \phi(\lambda)={\hatB}_\phi\Big(\phi(\lambda),\theta(\lambda),r(\lambda)\Big)\nonumber\\
r(\lambda)\partial_\lambda \theta(\lambda)={\hatB}_\theta\Big(\phi(\lambda),\theta(\lambda),r(\lambda)\Big)\nonumber\\
\partial_\lambda r(\lambda)={\hatB}_r\Big(\phi(\lambda),\theta(\lambda),r(\lambda)\Big)\ .
\ee
Here we used the fact that the magnetic field components are written in the spherical orthonormal basis $\hat \phi,\ \hat \theta,\ \hat r$, which correspond to longitude, latitude, and radius, respectively.

\subsection{Interpolation schemes}

\qsl allows one to use different interpolation schemes when calculating the values of the magnetic field vectors on the right-hand side of the above system of equations. This capability can be used to test the robustness of QSL maps on the interpolation order. The available interpolation schemes are
trilinear, triquadratic \citep{quadratic_interp} and tricubic \citep{cubic_interp}. 

To be able to write down the interpolation schemes explicitly, we need to introduce some notation first. The input magnetic field data cubes specify the values $\bm{B}^{\v}$ of the magnetic field at points with coordinates $\x_{\v}$, sampled on a rectilinear grid. Thus, $\v$ is a 3-vector, with each of its components running over the indices of the input 3d magnetic field array. Next, we would like to write down the value of $\hatbB$ at some arbitrary position $\x$. To do that, we need to identify the cell within the input array within which $\x$ lies. The interpolation kernels cover 8, 27, or 64 vertices neighboring that cell for trilinear, triquadratic and tricubic interpolation, respectively. Of the 8 vertices in the immediate neighborhood of $\x$, let us denote by $\c$ that vertex, which lies closest to the origin of the array. Therefore, for any of the above interpolation schemes, we can write:
\be\label{interp}
\hatbB(\x)=\sum\limits_{\v \in \hbox{stencil}}{\hatbB}^{\v+\c}f_{\v}\Big(\x-\x_{\c};\bm{h}_{\c}\Big)\ ,
\ee
where the sum runs over the 8, 27, or 64  vertices which span the respective interpolation stencil around $\x$. The physical dimensions of the array cell containing $\x$ are specified by the components of $\bm{h}_\c$. In other words, $\bm{h}_\c$ gives the span of that data cell in each dimension in the corresponding units (Mm or degrees, depending on geometry). The interpolation kernels $f_\v$ depend on the selected interpolation order. 

As an example, for trilinear interpolation, the interpolation kernels are given by:
\be
f_{0,0,0}(\x;\h)&=&\left(1-\frac{x_0}{h_0}\right)\left(1-\frac{x_1}{h_1}\right)\left(1-\frac{x_2}{h_2}\right)\\
f_{0,0,1}(\x;\h)&=&\left(1-\frac{x_0}{h_0}\right)\left(1-\frac{x_1}{h_1}\right)\left(\frac{x_2}{h_2}\right)\nonumber\\
f_{0,1,1}(\x;\h)&=&\left(1-\frac{x_0}{h_0}\right)\left(\frac{x_1}{h_1}\right)\left(\frac{x_2}{h_2}\right)\nonumber\\
&\cdots&\nonumber
\ee
When running \qsl for spherical geometry, the input magnetic field is sampled on a rectilinear grid in spherical coordinates. Thus, in the above equation, we have $x_0=\phi-\phi_\c$, $x_1=\theta-\theta_\c$, $x_2=r-r_\c$, while $h_0,\ h_1,\ h_2$ give the grid spacing in longitude, latitude and radius, respectively, for the cell containing $\x$.

\subsection{Field-line deviation using linearization}

The squashing factor $Q$ quantifies how neighboring field lines deviate from one another. The procedure described in \cite{Pariat12} for solving for $Q$ involves explicitly integrating three closely spaced field lines, after which one takes the finite differences in position of the footpoints of those field lines. Those differences in turn enter in the calculation of the squashing factor. However, calculating those field-line deviations using such a finite difference scheme puts severe constraints on the precision with which one should follow neighboring field lines. \cite{Pariat12}  quote a fractional precision of $10^{-8}$ for their calculation, which results in severe speed penalties. 

In \qsl, we alleviate that problem by calculating field-line deviations by linearizing the deviation equation as follows. The deviation between two neighboring field-lines $\x(\lambda)$ and $\x'(\lambda)$ is quantified by the difference in their positions: $\delta \x(\lambda)=\x'(\lambda)-\x(\lambda)$. Here we assume that when $\lambda=0$, the positions along the two field lines are infinitesimally apart. Thus, the field-line deviation can be calculated as follows:
\be\label{dev}
\partial_\lambda \delta \x(\lambda)=\hatbB\Big(\x'(\lambda)\Big)-\hatbB\Big(\x(\lambda)\Big)\approx \bigg(\delta  \x(\lambda)\cdot\bm{\nabla}_{\x} \bigg)\hatbB\Big(\x(\lambda)\Big)\ ,
\ee
where we used perturbation theory to linearize the equation in the second equality\footnote{A similar perturbative approach in studying certain analytic magnetic field configurations was used by \cite{2003ApJ...582.1172T}.}

In order to integrate the above equation, we need to be able to take the gradient of $\hatbB$. Recalling that $\hatbB(\x)$ is calculated using (\ref{interp}), taking the gradient is straightforward as it acts only on the interpolation kernels. In spherical coordinates, the field-line deviation is given explicitly below for reference\footnote{In eq.~(\ref{dev_sph}), we neglect terms that are suppressed by the ratio of the typical scale over which $\hatbB$ varies and the radius of the Sun. We will write down those terms explicitly elsewhere.}:
\be\label{dev_sph}
r(\lambda)\cos\Big(\theta(\lambda)\Big)\partial_\lambda \delta \phi(\lambda)&\approx&\sum\limits_{\v \in \hbox{stencil}}
{\hatB}^{\v+\c}_{\phi}\delta f_{\v}\Big(\delta\x(\lambda);\x(\lambda)-\x_\c;\h_\c\Big)
\nonumber\\
r(\lambda)\partial_\lambda \delta \theta(\lambda)&\approx&\sum\limits_{\v \in \hbox{stencil}}
{\hatB}^{\v+\c}_{\theta}\delta f_{\v}\Big(\delta\x(\lambda);\x(\lambda)-\x_\c;\h_\c\Big)
\nonumber\\
\partial_\lambda \delta r(\lambda)&\approx&\sum\limits_{\v \in \hbox{stencil}}
{\hatB}^{\v+\c}_r\delta f_{\v}\Big(\delta\x(\lambda);\x(\lambda)-\x_\c;\h_\c\Big)\ ,
\ee
where $\delta f$ is given by (after suppressing its arguments):
\be
\delta f_\v\equiv\bigg(\delta\phi(\lambda)\partial_\phi +\delta\theta(\lambda)\partial_\theta+\delta r(\lambda)\partial_r\bigg)f_{\v}\Big(\phi,\, \theta,\, r;\, \bm{h}_\c\Big)\Bigg|_{\phi=\phi(\lambda)-\phi_\c,\, \theta=\theta(\lambda)-\theta_\c,\, r=r(\lambda)-r_\c}\ ,
\ee
where the derivatives are taken analytically in the code for each interpolation kernel.

\subsection{Integration methods}

\qsl offers a choice between two integration schemes for integrating the field lines and field-line deviation vectors. One can use either an explicit Euler scheme, or an adaptive Runge-Kutta Cash-Karp method \citep{Cash:1990:VOR:79505.79507} provided by the Boost \verb|runge_kutta_cash_karp54| stepper algorithm. The latter method can easily be substituted with any of the other integration methods offered by the Boost library.

The field line deviation vectors $\delta \x$ needed for the squashing factor calculation, are solved by \qsl using eq.~(\ref{dev}) (or for spherical geometry, using (\ref{dev_sph})) instead of the finite difference scheme of \cite{Pariat12}. This allows us to relax the precision and accuracy tolerances by many orders of magnitude. For the adaptive stepper, we have found a value of $10^{-2}$ to be more than sufficient for the real world example explored in this paper (see below). This tolerance is six orders of magnitude larger than the one quoted by \cite{Pariat12}. Choosing an explicit fixed-step Euler scheme with roughly 5 samplings per grid spacing gives about an order of magnitude speed-up relative to the adaptive stepper implementation that is incorporated in \qsl. One trades accuracy for such a speed-up. However, in our experiments, we have not encountered cases where using the adaptive stepper was beneficial. We still consider the adaptive stepper useful as it can be used for testing the convergence properties of the Euler scheme for the particular problem at hand.

\subsection{Squashing factor calculation}

Having introduced the basic equations allowing us to integrate the magnetic field lines and field-line deviations, we proceed to describe the calculation of the squashing factor. To calculate the squashing factor, we use Method 3 of \cite{Pariat12}. That requires projecting the components ($\delta \x_\perp$) of $\delta \x$ perpendicular to the field line tangent, given by $\hatbB$. In spherical coordinates, we can do that if we write both $\hatbB$ and $\delta \x$ in the spherical orthonormal basis spanned by $\hat \phi,\ \hat\theta, \ \hat r$. The magnetic field is given by interpolating the input magnetic field using (\ref{interp}), while the components of the field line deviation vector are given by:
\be
\delta x_\phi=r(\lambda)\cos\Big(\theta(\lambda)\Big) \delta \phi(\lambda)\ , \ \ \ 
\delta x_\theta=r(\lambda) \delta \theta(\lambda)\ , \ \ \ 
\delta x_r=\delta r(\lambda)\ .
\ee
Since our choice of basis is orthonormal, calculating\footnote{Notice that normalizing a vector field and then interpolating it is not equivalent to interpolating a vector field and then normalizing it. It is up to the user to supply arrays of the magnetic field that are sampled finely enough to make this difference unimportant. To speed up the code, we first normalize the magnetic field data cubes, and only then interpolate them. Thus, $\hatbB(\x)$ as calculated from (\ref{interp}) is not guaranteed to be a unit vector. Therefore, when extracting $\delta \x_\perp$ the code explicitly normalizes $\hatbB(\x)$ beforehand.} $\delta \x_\perp=\delta \x-\hatbB\left(\hatbB\cdot\delta\x\right)$ numerically is straightforward. Here $\hatbB$ is evaluated at $\x(\lambda)$. 

To calculate $Q$ at a position $\x_0$, we need to integrate $\delta \x(\lambda)$ for two sets of initial conditions (denoted with superscripts):
\be\label{ic}
\delta \x^{(1)}(\lambda=0)= \hat\a\ , \ \mathrm{and} \ \ \ \delta \x^{(2)}(\lambda=0)= \hat\b\ ,
\ee
where $\lambda=0$ corresponds to the initial condition for the field line $\x(\lambda=0)=\x_0$, which passes through $\x_0$. The only restriction on vectors $\hat\a$ and $\hat\b$ is that they, combined with $\hatbB\Big(\x_0\Big)$, form an orthonormal basis at $\x_0$. Integrating the field line $\x(\lambda)$ along with $\delta \x^{(1)}(\lambda)$ and $\delta \x^{(2)}(\lambda)$ involves solving equations (\ref{fl}) and (\ref{dev}) (which in spherical coordinates, correspond to (\ref{fl_sph}) and (\ref{dev_sph})), forwards to $\lambda_\rF$, and then backwards to $\lambda_\rB$, subject to the initial conditions (\ref{ic}). Those values of $\lambda$ correspond to parameter values for which $\x(\lambda)$ reaches the boundary  of the region spanned by the input magnetic field (irrespective of which boundary: bottom, top or side). This definition is used for both open and closed field lines. A switch in \qsl allows one to identify open field lines and skip them from the QSL calculation.

As a side note, note that the standard squashing factor is calculated by knowing how much neighbouring field lines deviate from one field line \textit{footpoint} to the other. However, one can envision applications where more localized $Q$ values may also be of interest, especially in studies of reconnection  without natural boundaries, such as magnetospheric studies, or studies of CMEs, and especially interplanetary CMEs. Thus, \qsl allows calculating $Q$ by integrating field lines to $\lambda_F$ and $\lambda_B$ spanning a fixed maximum field line length. Note that choosing to enable that option gives a $Q$ which is no longer a global quantity, and is no longer constant along the length of a field line.

Moving on, let us define the solutions at the endpoints (irrespective of whether they are at the photosphere or not) as:
\be
\a_\rF\equiv\delta \x^{(1)}_\perp(\lambda_\rF)\ , \ \ \ \a_\rB\equiv\delta \x^{(1)}_\perp(\lambda_\rB)\ ,\nonumber\\
\b_\rF\equiv\delta \x^{(2)}_\perp(\lambda_\rF)\ , \ \ \ \b_\rB\equiv\delta \x^{(2)}_\perp(\lambda_\rB)\ .
\ee
With these definitions, after a bit of algebra, one can show that $Q$ as calculated using Method 3 of \cite{Pariat12} can be reduced to the following  expression\footnote{The easiest way to see that is to use the fact that $Q$ is covariant by construction (e.g. \cite{Pariat12}). Thus, one can pick one of the basis vectors at the location of $\x(\lambda_\rF)$ to be $\hat \a_\rF$. Then, the components of $\a_\rF$ in the plane perpendicular to the field line are ($a_\rF$,0), while those of $\b_\rF$ are $(\hat \a_\rF\cdot \b_\rF,b_\rF\sqrt{1- (\hat \a_\rF\cdot \hat \b_\rF)^2})$. One can write the analogous expressions for the location $\x(\lambda_\rB)$. One can then identify those vectors with the notation of \cite{Pariat12} by examining their Fig.~4. Thus, for example, one can identify $\a_\rF$ with their (dX2yc,dY2yc) and $\b_\rF$ with their (dX2xc,dY2xc). Using these identifications, after plugging in our components of $\a_\rF$, $\b_\rF$ and so on into their equation~(21), one recovers our equation (\ref{finalQ}).}, which is straightforward to implement numerically:
\be\label{finalQ}
Q=\frac{B_\rF B_\rB}{B^2_0}
\left[
a_\rF^2 b_\rB^2+
a_\rB^2 b_\rF^2-
2(\a_\rB\cdot \b_\rB)(\a_\rF\cdot \b_\rF)
\right]\ ,
\ee
where $B_\rF\equiv B(\x(\lambda_\rF))$, $B_\rB\equiv B(\x(\lambda_\rB))$ and $B_0\equiv B(\x(\lambda=0))$ give the unnormalized magnetic field magnitudes.

\newpage

\subsection{Adaptive refinements}

In order to be able to identify QSLs, one needs to resolve high-$Q$ regions, which correspond to thin surfaces, separating the quasi-connectivity domains  in 3D. Thus, a proper QSL code needs to perform adaptive refinements around those regions. One way to do that is to refine in regions where $Q$ (or its second derivative, for example) is larger than a predefined threshold.  In \qsl, we employ an alternative method, which identifies those domain boundaries much more robustly.  The method relies on using Field-line Length Edge (FLEDGE) maps which we introduce next.

\begin{figure}[t!]
\plotone{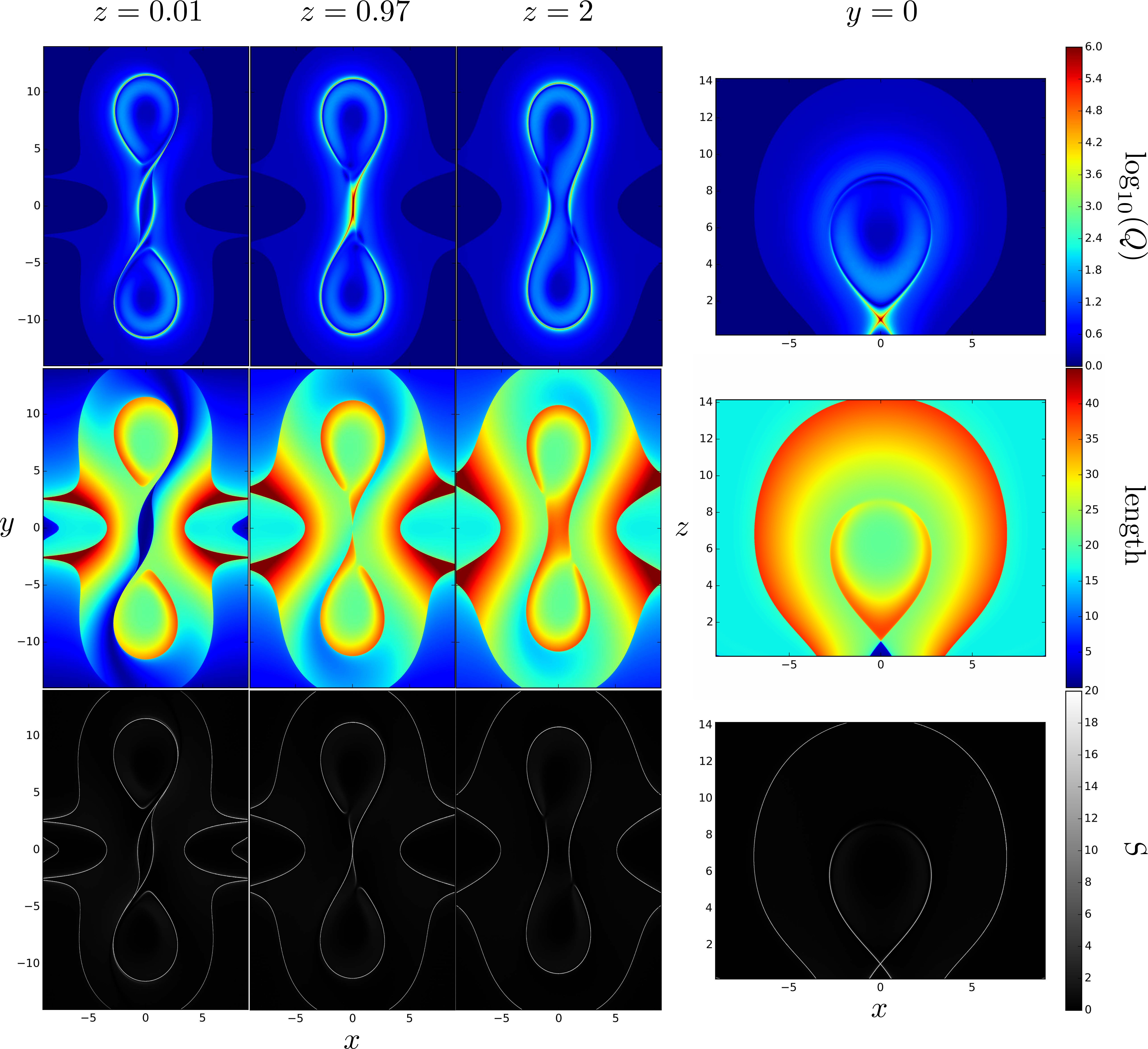}\caption{Horizontal and vertical planar slices of a TD flux rope. The first column shows horizontal sections taken below the HFT; the second column is a horizontal section through the HFT; and the third is taken above the HFT at the $z$ locations identified above each column ($x$, $y$ and $z$ are given in normalized coordinates). The rightmost column shows a vertical cross-section, similar to the one shown in Fig.\,\ref{3DTD}. The three rows correspond to: a QSL map, quantified by the logarithm of the squashing factor, $\log_{10}{Q}$; a field-line length (FLL) map (labelled by ``$\mathrm{length}$''); and a FLEDGE map (labelled by $S$), realized as the gradient magnitude from the Sobel operator applied to the FLL map. Note the correspondence between QSLs and large FLL jumps in the FLEDGE maps. Regions with open field lines (i.e field lines reaching either a top or later boundary of the computational box) around the flux rope are indicated with dark blue in the QSL map.}\label{TDslices}
\end{figure}

We define a FLEDGE map to be \textit{any} map of the changes of the length of neighbouring field lines\footnote{For an early paper exploring field-line length (FLL) discontinuities in 1D see \citep{Demoulin96a}. The authors there also suggest looking into the discontinuities in the footpoint distance as yet another indicator of connectivity boundaries. Here we focus solely on jumps in the FLLs.}. As an example, such changes can be mapped out using the gradient magnitude from the Sobel operator (i.e. a 2D or 3D gradient of FLL convolved with a simple smoothing function) applied to a 2D or 3D map of the length of field lines in a section or a volume. Examples of FLEDGE maps are shown in the last row of Figure~\ref{TDslices} (discussed further in the next section), where a Sobel filter was applied to the 2D maps of the Field-Line Length (FLL) shown in the middle row of that figure. 

Having introduced FLEDGE maps, let us move on to describe the way adaptive refinements are handled by \qsl. First, we sample the volume (or area, in the case of a slice output) of interest on a customizable rectilinear grid. Then that 2D/3D array is mapped to a 1D array using the following method.

One fills the slice or volume of interest with a Hilbert curve. This allows one to map the region of interest onto a one-dimensional curve. Thus, the array holding the $Q$ and FLL values in the code can be rendered one-dimensional, with successive elements of that array ordered according to position along the Hilbert curve. 

Hilbert curves carry the useful property that neighboring points on the Hilbert curve are necessarily close together in real space, although the opposite does not necessarily hold. In the spirit of the FLEDGE maps described above, our criterion for refining the sampling in a region of interest is checking whether the jump in the FLLs between two successive samples along the Hilbert curve surpasses a certain threshold. This threshold thus serves as a termination criterion of the adaptive refinements and is customizable in the code. For example, it can be easily modified to require a convergence in $Q$ value, instead.

One can envision many other possible choices, but we found the default convergence rule quite robust, converging on domain boundaries about an order of magnitude faster than using a threshold in $Q$ (or its second derivative along the Hilbert curve) as a refinement criterion. If the FLL jump threshold is surpassed, then the code calculates $Q$ (and the respective field-line length) halfway along the Hilbert curve between those two neighboring samples. 

For convenience, \qsl includes a code which takes the $Q$ (or FLL) array sampled along the Hilbert curve, and converts it into a 2D or 3D array of $Q$ values sampled on a rectilinear grid spanning the respective slice or volume of interest. If more than one $Q$ value is found in a cell around a grid point, the value the array converter assigns to that point is the maximum $Q$ value in that cell. If there are no samples in a grid cell, then the array converter interpolates the $\log(Q)$ values along the Hilbert curve to fill in the gap.

The Hilbert curve refinements can miss a point that lies in between samples that are not close along the Hilbert curve, but are close in real space. To alleviate that problem, after each refinement step, we shift the Hilbert curve by a small amount in real space; then reorder the $Q$ and FLL arrays along that new Hilbert curve; perform the refinement step again; and then shift back the Hilbert curve to its original position, reordering the arrays along that original curve. We have found that applying this shifting technique nearly eliminates such misses, and makes any artifacts irrelevant. 

The benefit of the Hilbert curve refinements (as opposed to using more sophisticated adaptive-mesh techniques) is that the code performing the adaptive refinements is about 50 lines long and requires no special book-keeping other than keeping track of the Hilbert coordinate of each point for which a $Q$ value is known. The functions responsible for the calculation of the $Q$ values are independent on the choice of refinement scheme. Thus, incorporating any other type of adaptive refinement in \qsl should be a straightforward coding exercise.

\section{Illustrative examples}\label{ex}

\subsection {Titov \& D{\'e}moulin flux rope}

\begin{figure}[t!]
\plotone{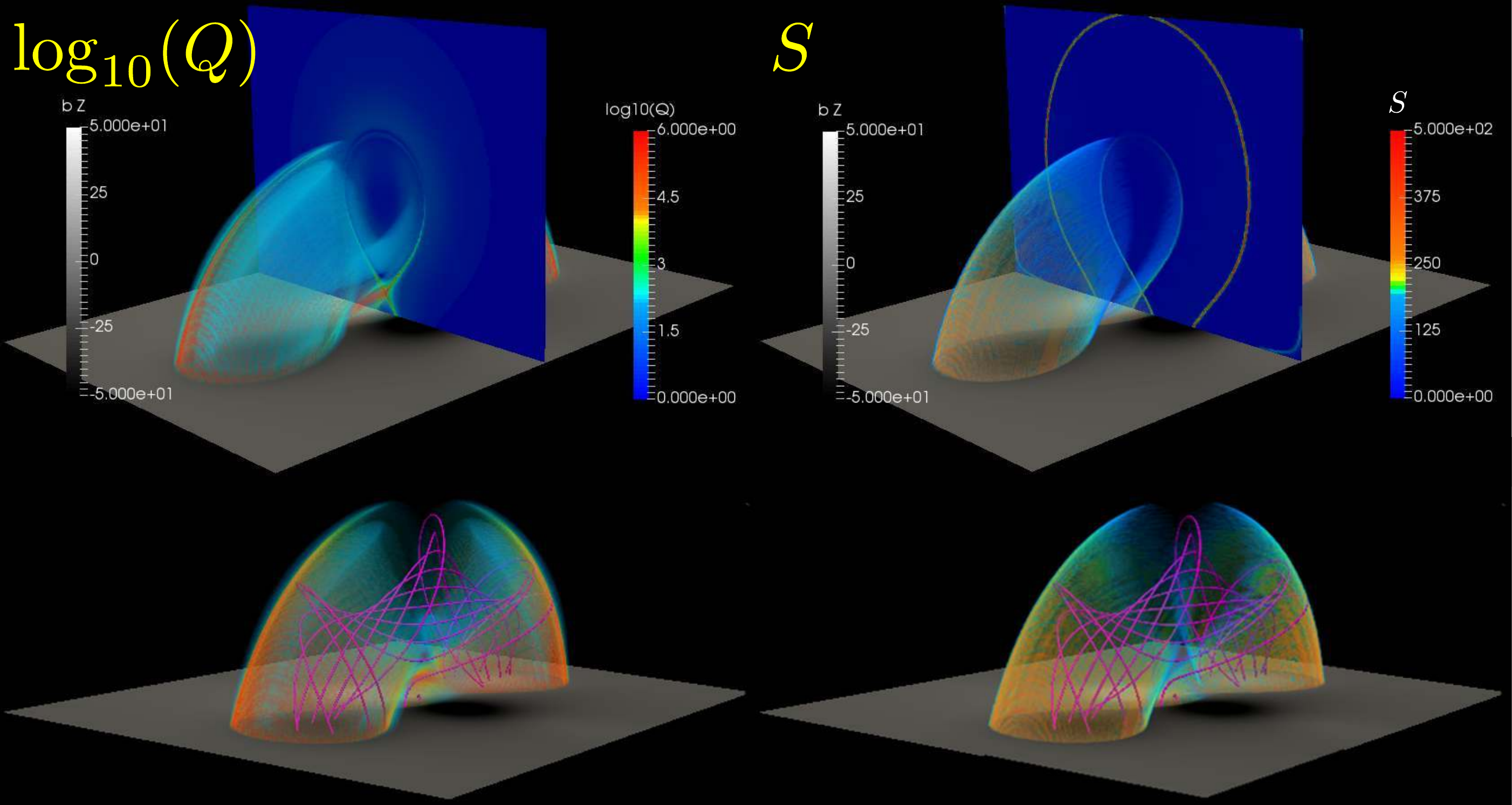}\caption{A 3D Cartesian rendering of the TD flux rope. The bottom grayscale section shows the vertical component of the  magnetic field. The left column shows a 3D QSL map (quantified by $Q$), while the right column shows the corresponding 3D FLEDGE map (quantified by the gradient magnitude from the 3D Sobel operator applied to the FLL map). A cut through the middle of the flux rope is shown as well highlighting the boundary of the flux rope and the HFT underneath, which is clearly visible in the QSL map, but not in the FLEDGE map.  The QSL associated with the transition to open field lines is made visible in the 2D FLEDGE section as well, although it is filtered out from the 3D rendering. Note the similarities between the QSL and FLEDGE maps. The striations seen in the 3D maps are an artefact of the rendering algorithm.}\label{3DTD}
\end{figure}

In order to illustrate the capabilities of \qsl, in this section we show several 2D and 3D QSL and FLEDGE maps obtained with the code. Originally, the theory of QSLs has been developed for the Cartesian analytical model of a flux rope following the construction of \citet[TD;][]{TD99}, which has served as an analytical case study for QSL calculation methods \citep{Pariat12}. Thus, the first results from \qsl we include are obtained for the numerical implementation of the TD flux rope model as given by \cite{vanBallegooijen08}. Those are shown in 2D in Fig.~\ref{TDslices} and in 3D in Fig.~\ref{3DTD}.

In Fig.\,\ref{TDslices}, we have shown horizontal slices through the FR at 3 different heights (first three columns), as well as vertical slices through the flux rope (fourth column). The three rows in the figure correspond to: a QSL map, quantified by the squashing factor, $Q$; a map of the field-line length(FLL); as well as a FLEDGE map, quantified by the gradient magnitude from a Sobel filter applied to the FLL map.

The height of the sections in the first column ($z=0.01$) is taken below the peak of the HFT, so that the QSLs have a 2J shape with the J's facing away from each other \citep[for a cartoon illustration see Fig.\,8 of][]{Savcheva12a, TD99}. For the chosen value of flux rope twist, the horizontal maps display QSL hooks that are almost closed on each other. The effect of twist on the hooks was first discussed in detail by \cite{Demoulin96b}, in particular how the flux-rope binding QSL wraps on itseld as the twist increases, and later was shown in observed fields by \cite{Savcheva12b} and \cite{Zhao16}.

The second column ($z=0.97$) shows a single S-shaped QSL because the cut passes through the HFT \citep[see Fig.\,8][]{Savcheva12a}. The third column ($z=2$) shows the QSL that encircles the flux rope when the cut is taken above the HFT. The curve is almost closed due to the large amount of twist in the rope.  The vertical section clearly shows the HFT under the flux rope core as the location where the QSLs that wraps around the flux rope intersects with itself.

From Fig.\,\ref{TDslices}, one can see that large values in the FLEDGE maps (locations where FLL jumps) correspond to quasi-domain boundaries, characterized by QSLs. Note that the FLEDGE maps also capture the boundary between open and closed field lines. The close correspondence between the QSL and FLEDGE maps is investigated further below.

\begin{figure}[t!]
\plotone{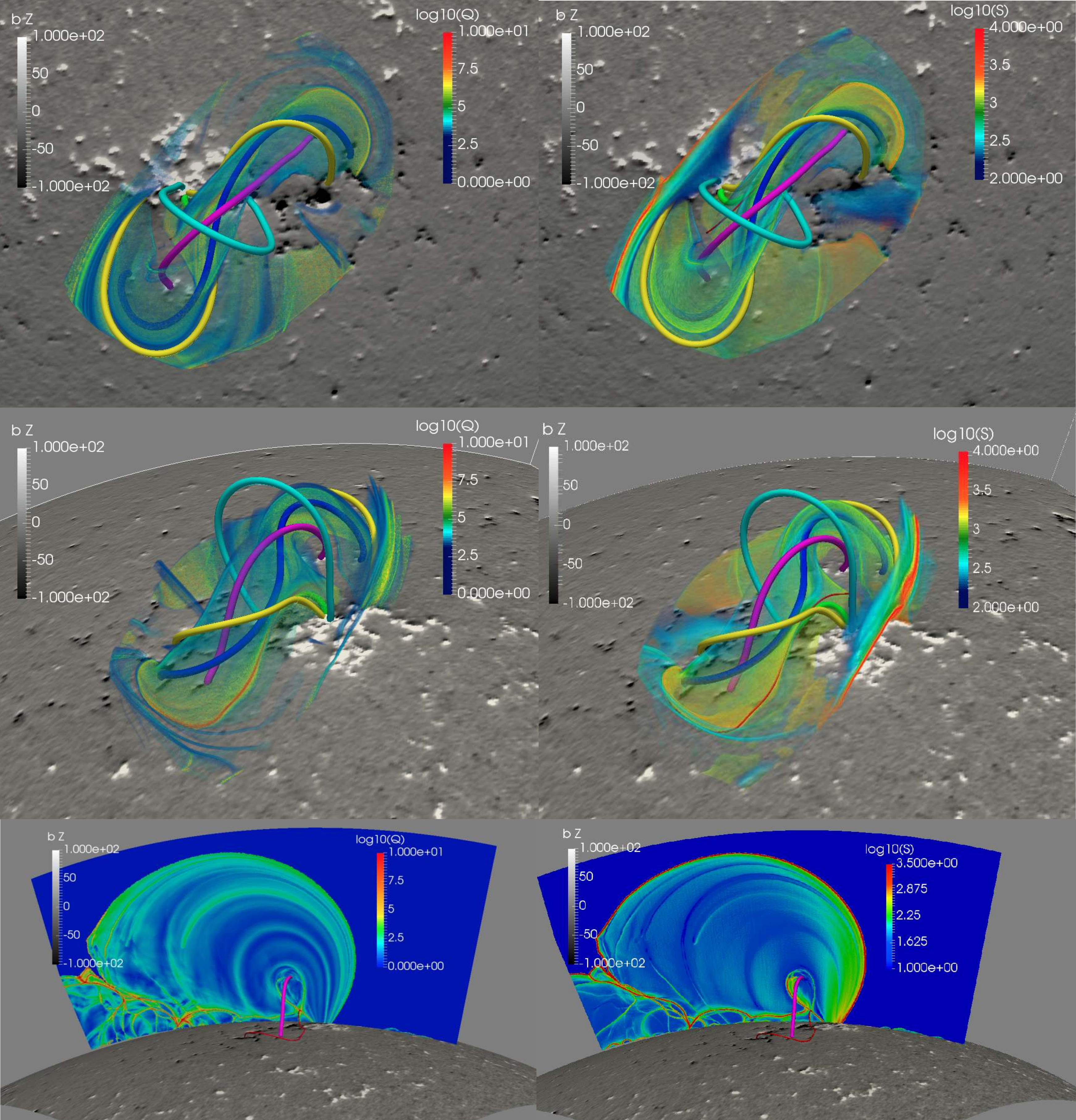}\caption{Shown are different views of the 3D QSL (left column) and FLEDGE (right column) maps for region  SOL2010-04-08. A fly-by movie around the 3D QSL structures shown here are available in the online version. The geometry of the modelled region is that of a spherical wedge. The grayscale spherical slice shows the HMI magnetogram used in generating the NLFFF model of the region. The field lines show: the core of the flux rope (magenta); the overlying ``potential'' arcade (cyan); 2J field lines (yellow); S-shaped field line (blue); the flare arcade (green, at low height in the center); and the HFT (thin red line). The 3D-rendered surface of $Q$ (better visible in the animated version of the figure) and the Sobel gradient magnitude outlines the cavity of the flux rope. In the bottom row we show a vertical planar section through the spherical domain in the middle of the flux rope. An animated version of this figure can be found at \url{https://bitbucket.org/tassev/qsl_squasher/downloads/3D_QSL.m4v}.}\label{100408slice}
\end{figure}

In  Fig.\,\ref{3DTD}, the 3D QSL (left column) and FLEDGE (right column) maps are rendered using Paraview. One can clearly see the QSL surface that wraps around the flux rope. In the top panel, a 2D section is added in the center of the rope showing the outline of the rope and the crossing of the QSL with itself at the HFT (the reddest part of the volume). A semi-transparent reddish feature from the 3D rendering can be seen to pass through the saddle point of the HFT in the 2D map. Note that 2D sections can be computed separately by \qsl, or can be extracted from the 3D volume of $Q$ in a visualization software. In the bottom panel of the Fig.\,\ref{3DTD} we have shown some sample field lines that belong to the flux rope and are contained within the 3D-rendered surface of the TD flux rope QSL. As in  Fig.\,\ref{TDslices}, in Fig.\,\ref{3DTD} one can clearly see the close resemblance between the QSL and FLEDGE maps in 3D. The most apparent difference is the fact that the HFT is not prominent in the FLEDGE map. Thus, while we can use the locations of the largest $Q$ values as a proxy for the location of the HFT, there is no such correspondence between values in the FLEDGE map and HFT's.

\begin{figure}[t!]
\plotone{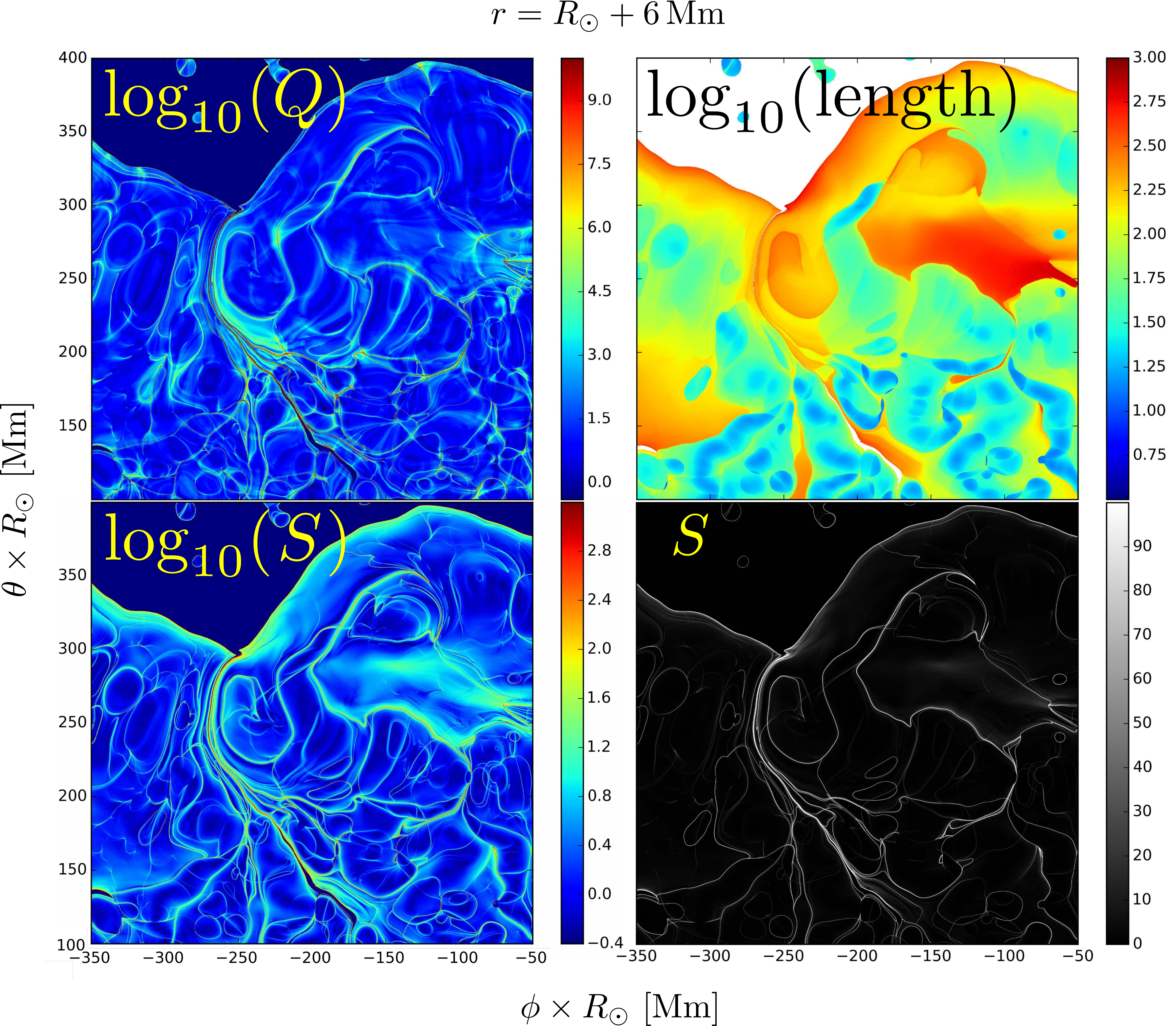}\caption{The figure shows slices at 6\,Mm above the photosphere. The top-left panel shows the squashing factor, $Q$; the top-right panel shows a map of the field-line length (FLL); while the bottom two panels show (two different scalings of) FLEDGE maps of the region as quantified by the gradient magnitude from the Sobel operator applied to the FLL map. The dark-blue region in the QSL map corresponds to open field lines. Note the qualitative match between the QSL and FLEDGE maps for this real-world example.}\label{1004083D}
\end{figure}

\subsection {SOL2010-04-08 sigmoidal region}

Next, we demonstrate the capabilities of \qsl for a data-constrained unstable magnetic field model in spherical coordinates, produced with the flux rope insertion method \citep{vanBallegooijen04,Savcheva09}. The model is of SOL2010-04-08 sigmoidal region that produced a B-class flare and a CME on 08 April 2010 and its stability has been studied in detail in \cite{Su11}. The unstable model analyzed here is produced by addition of axial flux to the best-fit marginally stable model, so that a residual Lorentz force exists, which prevents the field from reaching a non-linear force-free equilibrium during the magnetofrictional relaxation of the field \citep{Savcheva15}. Such an unstable model has been used by \cite{Kliem13} and \cite{Savcheva17} to produce an MHD eruption from this region. The lower boundary for the QSL calculation is set at 2\,Mm since lower than that the magnetic field models contain many 
{\bf low-lying and small scale NPs, separatrices, and QSLs assosciated with bald patches \citep[BPs;][]{Titov93}, which contain infinite values of $Q$, yet the field lines that pass through these features do not propagate to large heights.} The introduction of that boundary significantly speeds up the computation as discussed in \cite{Savcheva12b}. 

Our 3D results for  SOL2010-04-08 are shown in Fig.~\ref{100408slice}. The left column shows the QSL map of the region, while the right column, the FLEDGE map, computed in the same way as for the TD flux rope. 

The first two rows of Fig.~\ref{100408slice} show a top and side view of the spherical wedge domain of the magnetic field computation, including field lines sampling different quasi-connectivity domains, including the core of the flux rope (magenta) and the overlying arcade (cyan). The bottom row shows a planar vertical slice through the spherical domain in the middle of the flux rope. One can see the outline of the flux rope and the HFT (red field line) underneath. Notice that the HFT has already reached a significant height in the shown iteration of the magnetofrictional evolution \citep[see Fig.~10 in][]{Savcheva16a} and the erupting flux rope is in the process to turn into a CME. 

The 3D renderings in Fig.~\ref{100408slice} show QSLs associated with the flux rope as well as with the open field lines (seen better in the 2D slices in the bottom row of Fig.~\ref{100408slice}). Due to the complexity of the figure, we recommend one to view the animated version.

As was the case with the TD flux rope, in this realistic example, the FLEDGE map recovers the general flux rope structure seen in the  QSL map, including the intersection of the QSL surfaces at the location of the HFT. Yet, the HFT itself is not readily identified in the FLEDGE map.

However, calculating the 3D FLEDGE maps for SOL2010-04-08 took about 3\,min on a consumer workstation GPU (AMD W8100). The calculation of the 3D QSL map of the same region took 2 orders of magnitude more time because of the need for adaptive refinements\footnote{As a reference, without the adaptive refinements, for the same number of samples, the QSL calculation proceeds about twice slower than the FLEDGE map calculation.}. The main reason for this enormous difference is the fact that one does not need to perform adaptive refinements when computing FLEDGE maps as jumps in FLL are readily identified (see top-right panel of Fig.~\ref{1004083D}) even at low resolution, unlike local spikes in $Q$. Thus, we argue that FLEDGE maps offer a computationally cheap substitute of QSL maps that can be especially useful in the preliminary stages of any (quasi-)topological studies.

In Fig.\,\ref{1004083D} we have shown a spherical surface slice  below the peak of the HFT. This can be recognized by the QSL pattern around coordinates (-200\,Mm,\,300\,Mm) in the top-left panel, showing the 2D QSL map. That QSL pattern can be recognized as the 2J-pattern seen in the first column of Fig.~\ref{TDslices}, as well as in the cartoon shown in Fig.~8 of \cite{Savcheva12a}. The QSL map is certainly complicated due to the intrinsic complexity of the observed HMI magnetic field (no smoothing has been applied). This effect of real magnetic fields on the complexity of QSL maps has been discussed in detail in \cite{Savcheva12a,Savcheva12b}.  Yet, from the vertical slice in  Fig.~\ref{100408slice}, one can see that most of the complex structures are contained at low heights above the photosphere, and do not interfere substantially with one's ability to read the flux-rope structure from the 3D QSL maps, thus highlighting the importance of calculating QSL maps in three dimensions.

For comparison, in the other three panels of Fig.\,\ref{1004083D}, we show the corresponding FLL map (top-right), as well as the FLEDGE map with two different scalings and color-codings in the bottom two panels. Note that both the QSL and FLEDGE maps for the most part identify the same domain boundaries. However, we leave quantitative comparison between QSLs and FLEDGE maps for future work as one can envision special cases when a QSL is present without an associated major jump in field-line length.

\section{Summary}\label{summary}

In this paper we presented \qsl: a free, publicly available, open-source code for fast calculation of Quasi-Separatrix Layer maps in two or three dimensions. It requires an input magnetic field sampled on a rectilinear grid in Cartesian or spherical coordinates.

We benchmarked the code by calculating 3D QSL maps for a model of the SOL2010-04-08 sigmoidal region on a consumer workstation GPU (AMD W8100). We found that the code achieves large processing speeds for three main reasons, each of which results in an order-of-magnitude speed-up:
\begin{itemize}
\item Running the code on the GPU as opposed to the workstation CPU results in about an order of magnitude speed-up. 
\item Compared to previous studies \citep[e.g. ][]{Pariat12}, we drastically relax the precision requirements for the QSL calculation. We do that by applying perturbation theory when calculating field-line deviations, which are necessary for calculating the squashing factor, quantifying the QSL strength. 
\item We use a new boundary detection criterion between quasi-connectivity domains, which quickly identifies possible QSL locations which need to be finely sampled by the code. That boundary detection criterion relies on finding the locations of abrupt field-line length changes. A map of these jumps in field-line length we dub a FLEDGE map. We find that using such FLL jumps as a refinement criterion, instead of a threshold in $Q$ (or its second derivative), results in an order of magnitude speed-up of the code.
\end{itemize}
For the realistic model discussed above, we clocked \qsl at several million $Q$ values per minute, which implies that a representative 3D QSL map can be obtained within a few hours.

We also presented a quick-and-dirty alternative to QSL maps: FLEDGE maps, which can be optionally output by \qsl. We show that, for the most part, FLEDGE maps and QSL maps identify similar topological features. Constructing high-resolution 3D FLEDGE maps with \qsl can be completed in minutes -- two orders of magnitude faster than calculating the corresponding 3D QSL maps. The main reason for this difference is the fact that one does not need to perform adaptive refinements when computing FLEDGE maps as jumps in field-line length are readily identified even at low resolution, unlike local spikes in $Q$. Thus, we argue that FLEDGE maps offer a computationally cheap substitute of QSL maps that can be especially useful in the preliminary stages of any (quasi-)topological studies.

The potential advantages to the solar physics community of having such freely-available, open-source codes are largely unexplored beyond published data-reduction pipelines. One of our goals in making \qsl public is stimulating others to get involved in a collaborative effort to produce codes open to inspection and verification. This has the benefit of avoiding the duplication of coding efforts and waste of public resources, as well as decoupling the scientific and coding efforts. 

\qsl can be found at \url{https://bitbucket.org/tassev/qsl_squasher/}.

\acknowledgments

{\it Hinode} is a Japanese mission developed, launched, and operated by ISAS/JAXA in partnership with NAOJ, NASA, and STFC (UK). Additional operational support is provided by ESA, NSC (Norway). This work was supported by NASA contract NNM07AB07C to SAO. This work was primarily supported by the Air Force Office of Scientific Research under award FA9550-15-1-0030 to UCAR and subaward to SAO Z15-12504 AFGL. The QSL computations of the April 2010 region was partially supported by NASA HSR grant to SAO NNX16AH87G. We would like to thank Etienne Pariat, Pascal D{\'e}moulin, and Vlacheslav S. Titov for useful suggestions on the preprint version. We also would like Edward DeLuca for useful discussions throughout the writing of the code and the paper.    

\software{C++ Boost library \url{http://www.boost.org/}, C++ VexCL library \url{https://github.com/ddemidov/vexcl}, \citep{vexcl}; OpenCL \url{https://www.khronos.org/opencl/}; Python SciPy \citep{scipy}, \url{https://www.scipy.org/}, PyEVTK \url{https://www.python.org/}, \url{https://bitbucket.org/pauloh/pyevtk}; POCL \url{http://portablecl.org/}, \citep{pocl}; Paraview \url{http://www.paraview.org/}; VisIt \url{https://wci.llnl.gov/simulation/computer-codes/visit/}; Mayavi \url{http://docs.enthought.com/mayavi/mayavi/}, using VTK format \url{http://www.vtk.org/}}

\bibliographystyle{apj}  
\bibliography{Antonia_sig_topo_v3}       

\begin{thebibliography}{67}
\expandafter\ifx\csname natexlab\endcsname\relax\def\natexlab#1{#1}\fi

\bibitem[{{Abramenko} \& {Yurchishin}(1996)}]{Abramenko96}
{Abramenko}, V.~I. \& {Yurchishin}, V.~B. 1996, \solphys, 168, 47

\bibitem[{{Antiochos} {et~al.}(2011){Antiochos}, {Miki{\'c}}, {Titov},
  {Lionello}, \& {Linker}}]{Antiochos11}
{Antiochos}, S.~K., {Miki{\'c}}, Z., {Titov}, V.~S., {Lionello}, R., \&
  {Linker}, J.~A. 2011, \apj, 731, 112

\bibitem[{{Aulanier} {et~al.}(2012){Aulanier}, {Janvier}, \&
  {Schmieder}}]{Aulanier12}
{Aulanier}, G., {Janvier}, M., \& {Schmieder}, B. 2012, \aap, 543, A110

\bibitem[{Aulanier {et~al.}(2005)Aulanier, Pariat, \&
  D{\'e}moulin}]{Aulanier05b}
Aulanier, G., Pariat, E., \& D{\'e}moulin, P. 2005, Astronomy and Astrophysics,
  444, 961

\bibitem[{Aulanier {et~al.}(2006)Aulanier, Pariat, D{\'e}moulin, \&
  DeVore}]{Aulanier06b}
Aulanier, G., Pariat, E., D{\'e}moulin, P., \& DeVore, C.~R. 2006, Solar
  Physics, 238, 347

\bibitem[{Aulanier {et~al.}(2010)Aulanier, T{\"o}r{\"o}k, D{\'e}moulin, \&
  DeLuca}]{Aulanier10}
Aulanier, G., T{\"o}r{\"o}k, T., D{\'e}moulin, P., \& DeLuca, E.~E. 2010, The
  Astrophysical Journal, 708, 314

\bibitem[{Baker {et~al.}(2009)Baker, van Driel-Gesztelyi, Mandrini,
  D{\'e}moulin, \& Murray}]{Baker09}
Baker, D., van Driel-Gesztelyi, L., Mandrini, C.~H., D{\'e}moulin, P., \&
  Murray, M.~J. 2009, The Astrophysical Journal, 705, 926

\bibitem[{Cash \& Karp(1990)}]{Cash:1990:VOR:79505.79507}
Cash, J.~R. \& Karp, A.~H. 1990, ACM Trans. Math. Softw., 16, 201

\bibitem[{{Chandra} {et~al.}(2009){Chandra}, {Schmieder}, {Aulanier}, \&
  {Malherbe}}]{Chandra09}
{Chandra}, R., {Schmieder}, B., {Aulanier}, G., \& {Malherbe}, J.~M. 2009,
  \solphys, 258, 53

\bibitem[{{Chintzoglou} {et~al.}(2017){Chintzoglou}, {Vourlidas}, {Savcheva},
  {Tassev}, {Beltran}, \& {Stenborg}}]{Chintzoglou16}
{Chintzoglou}, G., {Vourlidas}, A., {Savcheva}, A., {Tassev}, S., {Beltran},
  S., \& {Stenborg}, G. 2017, \apj, submitted

\bibitem[{{Demidov} {et~al.}(2012){Demidov}, {Ahnert}, {Rupp}, \&
  {Gottschling}}]{vexcl}
{Demidov}, D., {Ahnert}, K., {Rupp}, K., \& {Gottschling}, P. 2012, ArXiv
  e-prints

\bibitem[{D{\'e}moulin {et~al.}(1994)D{\'e}moulin, H{\'e}noux, \&
  Mandrini}]{Demoulin94}
D{\'e}moulin, P., H{\'e}noux, J.~C., \& Mandrini, C.~H. 1994, Astronomy and
  Astrophysics, 285, 1023

\bibitem[{D{\'e}moulin {et~al.}(1996{\natexlab{a}})D{\'e}moulin, H{\'e}noux,
  Priest, \& Mandrini}]{Demoulin96a}
D{\'e}moulin, P., H{\'e}noux, J.~C., Priest, E.~R., \& Mandrini, C.~H.
  1996{\natexlab{a}}, Astronomy and Astrophysics, 308, 643

\bibitem[{{D{\'e}moulin} \& {Priest}(1997)}]{DemoulinP97}
{D{\'e}moulin}, P. \& {Priest}, E.~R. 1997, \solphys, 175, 123

\bibitem[{D{\'e}moulin {et~al.}(1996{\natexlab{b}})D{\'e}moulin, Priest, \&
  Lonie}]{Demoulin96b}
D{\'e}moulin, P., Priest, E.~R., \& Lonie, D.~P. 1996{\natexlab{b}}, Journal of
  Geophysical Research, 101, 7631

\bibitem[{Dodgson(1997)}]{quadratic_interp}
Dodgson, N.~A. 1997, IEEE Trans. on Image Processing, 1322

\bibitem[{{Downs} {et~al.}(2016){Downs}, {Titov}, {Qiu}, {T{\"o}r{\"o}k},
  {Linker}, \& {Mikic}}]{Downs16}
{Downs}, C., {Titov}, V., {Qiu}, J., {T{\"o}r{\"o}k}, T., {Linker}, J., \&
  {Mikic}, Z. 2016, SHINE Conference 2016

\bibitem[{{Gorbachev} \& {Somov}(1988)}]{GorbatchevSomov88}
{Gorbachev}, V.~S. \& {Somov}, B.~V. 1988, \solphys, 117, 77

\bibitem[{{Inoue} {et~al.}(2012){Inoue}, {Magara}, {Watari}, \&
  {Choe}}]{Inoue12}
{Inoue}, S., {Magara}, T., {Watari}, S., \& {Choe}, G.~S. 2012, \apj, 747, 65

\bibitem[{J\"a\"askel\"ainen {et~al.}(2015)J\"a\"askel\"ainen, de~La~Lama,
  Schnetter, Raiskila, Takala, \& Berg}]{pocl}
J\"a\"askel\"ainen, P., de~La~Lama, C.~S., Schnetter, E., Raiskila, K., Takala,
  J., \& Berg, H. 2015, International Journal of Parallel Programming, 43, 752

\bibitem[{{Janvier} {et~al.}(2014){Janvier}, {Aulanier}, {Bommier},
  {Schmieder}, {D{\'e}moulin}, \& {Pariat}}]{Janvier14}
{Janvier}, M., {Aulanier}, G., {Bommier}, V., {Schmieder}, B., {D{\'e}moulin},
  P., \& {Pariat}, E. 2014, \apj, 788, 60

\bibitem[{{Janvier} {et~al.}(2013){Janvier}, {Aulanier}, {Pariat}, \&
  {D{\'e}moulin}}]{Janvier13}
{Janvier}, M., {Aulanier}, G., {Pariat}, E., \& {D{\'e}moulin}, P. 2013, \aap,
  555, A77

\bibitem[{{Janvier} {et~al.}(2016){Janvier}, {Savcheva}, {Pariat}, {Tassev},
  {Millholland}, {Bommier}, {McCauley}, {McKillop}, \& {Dougan}}]{Janvier16}
{Janvier}, M., {Savcheva}, A., {Pariat}, E., {Tassev}, S., {Millholland}, S.,
  {Bommier}, V., {McCauley}, P., {McKillop}, S., \& {Dougan}, F. 2016, \aap,
  591, A141

\bibitem[{{Jiang} \& {Feng}(2012{\natexlab{a}})}]{Jiang12b}
{Jiang}, C. \& {Feng}, X. 2012{\natexlab{a}}, \apj, 749, 135

\bibitem[{{Jiang} \& {Feng}(2012{\natexlab{b}})}]{Jiang12a}
---. 2012{\natexlab{b}}, \solphys, 281, 621

\bibitem[{{Jiang} {et~al.}(2016){Jiang}, {Wu}, {Feng}, \& {Hu}}]{Jiang16}
{Jiang}, H., {Wu}, S.~T., {Feng}, X., \& {Hu}, Q. 2016, Nat. Com., 682, 11152

\bibitem[{Jones {et~al.}(2001)Jones, Oliphant, Peterson, {et~al.}}]{scipy}
Jones, E., Oliphant, T., Peterson, P., {et~al.} 2001, {SciPy}: Open source
  scientific tools for {Python}, [Online; accessed 2016-06-27]

\bibitem[{Keys(1981)}]{cubic_interp}
Keys, R.~G. 1981, IEEE Trans. Acoust., Speech, Signal Process, 1153

\bibitem[{{Kliem} {et~al.}(2013){Kliem}, {Su}, {van Ballegooijen}, \&
  {DeLuca}}]{Kliem13}
{Kliem}, B., {Su}, Y.~N., {van Ballegooijen}, A.~A., \& {DeLuca}, E.~E. 2013,
  \apj, 779, 129

\bibitem[{Lawrence \& Gekelman(2009)}]{PhysRevLett.103.105002}
Lawrence, E.~E. \& Gekelman, W. 2009, Phys. Rev. Lett., 103, 105002

\bibitem[{{Linker} {et~al.}(2011){Linker}, {Lionello}, {Miki{\'c}}, {Titov}, \&
  {Antiochos}}]{Linker11}
{Linker}, J.~A., {Lionello}, R., {Miki{\'c}}, Z., {Titov}, V.~S., \&
  {Antiochos}, S.~K. 2011, \apj, 731, 110

\bibitem[{{Liu} {et~al.}(2014){Liu}, {Titov}, {Gou}, {Wang}, {Liu}, \&
  {Wang}}]{Liu14}
{Liu}, R., {Titov}, V.~S., {Gou}, T., {Wang}, Y., {Liu}, K., \& {Wang}, H.
  2014, \apj, 790, 8

\bibitem[{{Longcope} \& {Strauss}(1994)}]{1994ApJ...437..851L}
{Longcope}, D.~W. \& {Strauss}, H.~R. 1994, \apj, 437, 851

\bibitem[{{Malanushenko} {et~al.}(2012){Malanushenko}, {Schrijver}, {DeRosa},
  {Wheatland}, \& {Gilchrist}}]{Malanushenko12}
{Malanushenko}, A., {Schrijver}, C.~J., {DeRosa}, M.~L., {Wheatland}, M.~S., \&
  {Gilchrist}, S.~A. 2012, \apj, 756, 153

\bibitem[{{Masson} {et~al.}(2013){Masson}, {Antiochos}, \& {DeVore}}]{Masson13}
{Masson}, S., {Antiochos}, S.~K., \& {DeVore}, C.~R. 2013, \apj, 771, 82

\bibitem[{{Pariat} \& {D{\'e}moulin}(2012)}]{Pariat12}
{Pariat}, E. \& {D{\'e}moulin}, P. 2012, \aap, 541, A78

\bibitem[{{Parnell} {et~al.}(2010){Parnell}, {Maclean}, \&
  {Haynes}}]{Parnell10}
{Parnell}, C.~E., {Maclean}, R.~C., \& {Haynes}, A.~L. 2010, \apjl, 725, L214

\bibitem[{{Pontin} {et~al.}(2016){Pontin}, {Galsgaard}, \&
  {D{\'e}moulin}}]{Pontin16}
{Pontin}, D., {Galsgaard}, K., \& {D{\'e}moulin}, P. 2016, \solphys, 291, 1739

\bibitem[{{Pontin} \& {Hornig}(2015)}]{Pontin15}
{Pontin}, D.~I. \& {Hornig}, G. 2015, \apj, 805, 47

\bibitem[{Priest \& D{\'e}moulin(1995)}]{Priest95}
Priest, E.~R. \& D{\'e}moulin, P. 1995, Journal of Geophysical Research, 100,
  23443

\bibitem[{{Savcheva} {et~al.}(2017){Savcheva}, {Lugaz}, {van der Holst},
  {Evans}, {Zhang}, {DeLuca}, \& {Reeves}}]{Savcheva17}
{Savcheva}, A., {Lugaz}, N., {van der Holst}, B., {Evans}, R., {Zhang}, Z.,
  {DeLuca}, E.~E., \& {Reeves}, K. 2017, Nature, in prep

\bibitem[{{Savcheva} {et~al.}(2016){Savcheva}, {Pariat}, {McKillop},
  {McCauley}, {Hanson}, {Su}, \& {DeLuca}}]{Savcheva16a}
{Savcheva}, A., {Pariat}, E., {McKillop}, S., {McCauley}, P., {Hanson}, E.,
  {Su}, Y., \& {DeLuca}, E.~E. 2016, \apj, 817, 43

\bibitem[{{Savcheva} {et~al.}(2015){Savcheva}, {Pariat}, {McKillop},
  {McCauley}, {Hanson}, {Su}, {Werner}, \& {DeLuca}}]{Savcheva15}
{Savcheva}, A., {Pariat}, E., {McKillop}, S., {McCauley}, P., {Hanson}, E.,
  {Su}, Y., {Werner}, E., \& {DeLuca}, E.~E. 2015, \apj, 810, 96

\bibitem[{{Savcheva} {et~al.}(2012b){Savcheva}, {Pariat}, {van Ballegooijen},
  {Aulanier}, \& {DeLuca}}]{Savcheva12b}
{Savcheva}, A., {Pariat}, E., {van Ballegooijen}, A., {Aulanier}, G., \&
  {DeLuca}, E. 2012b, \apj, 750, 15, S12

\bibitem[{Savcheva \& van Ballegooijen(2009)}]{Savcheva09}
Savcheva, A. \& van Ballegooijen, A. 2009, The Astrophysical Journal, 703, 1766

\bibitem[{{Savcheva} {et~al.}(2012a){Savcheva}, {van Ballegooijen}, \&
  {DeLuca}}]{Savcheva12a}
{Savcheva}, A.~S., {van Ballegooijen}, A.~A., \& {DeLuca}, E.~E. 2012a, \apj,
  744, 78

\bibitem[{Schrijver {et~al.}(2011)Schrijver, Aulanier, Title, Pariat, \&
  Delann{\'e}e}]{Schrijver11}
Schrijver, C.~J., Aulanier, G., Title, A.~M., Pariat, E., \& Delann{\'e}e, C.
  2011, The Astrophysical Journal, 738, 167

\bibitem[{{Schrijver} {et~al.}(2006){Schrijver}, {De Rosa}, {Metcalf}, {Liu},
  {McTiernan}, {R{\'e}gnier}, {Valori}, {Wheatland}, \&
  {Wiegelmann}}]{Schrijver06}
{Schrijver}, C.~J., {De Rosa}, M.~L., {Metcalf}, T.~R., {Liu}, Y., {McTiernan},
  J., {R{\'e}gnier}, S., {Valori}, G., {Wheatland}, M.~S., \& {Wiegelmann}, T.
  2006, \solphys, 235, 161

\bibitem[{{Schrijver} {et~al.}(2010){Schrijver}, {DeRosa}, \&
  {Title}}]{Schrijver10}
{Schrijver}, C.~J., {DeRosa}, M.~L., \& {Title}, A.~M. 2010, \apj, 719, 1083

\bibitem[{Su {et~al.}(2011)Su, Surges, van Ballegooijen, DeLuca, \&
  Golub}]{Su11}
Su, Y., Surges, V., van Ballegooijen, A., DeLuca, E.~E., \& Golub, L. 2011, The
  Astrophysical Journal, 734, 53

\bibitem[{Titov(2007)}]{Titov07}
Titov, V.~S. 2007, The Astrophysical Journal, 660, 863

\bibitem[{Titov \& D{\'e}moulin(1999)}]{TD99}
Titov, V.~S. \& D{\'e}moulin, P. 1999, Astronomy and Astrophysics, 351, 707

\bibitem[{{Titov} {et~al.}(2003){Titov}, {Galsgaard}, \&
  {Neukirch}}]{2003ApJ...582.1172T}
{Titov}, V.~S., {Galsgaard}, K., \& {Neukirch}, T. 2003, \apj, 582, 1172

\bibitem[{Titov {et~al.}(2002)Titov, Hornig, \& D{\'e}moulin}]{Titov02}
Titov, V.~S., Hornig, G., \& D{\'e}moulin, P. 2002, Journal of Geophysical
  Research, 107, 1164

\bibitem[{{Titov} {et~al.}(2011){Titov}, {Miki{\'c}}, {Linker}, {Lionello}, \&
  {Antiochos}}]{Titov11}
{Titov}, V.~S., {Miki{\'c}}, Z., {Linker}, J.~A., {Lionello}, R., \&
  {Antiochos}, S.~K. 2011, \apj, 731, 111

\bibitem[{{Titov} {et~al.}(2012){Titov}, {Mikic}, {T{\"o}r{\"o}k}, {Linker}, \&
  {Panasenco}}]{Titov12}
{Titov}, V.~S., {Mikic}, Z., {T{\"o}r{\"o}k}, T., {Linker}, J.~A., \&
  {Panasenco}, O. 2012, \apj, 759, 70

\bibitem[{Titov {et~al.}(1993)Titov, Priest, \& D{\'e}moulin}]{Titov93}
Titov, V.~S., Priest, E.~R., \& D{\'e}moulin, P. 1993, Astronomy and
  Astrophysics, 276, 564

\bibitem[{{T{\"o}r{\"o}k} {et~al.}(2011){T{\"o}r{\"o}k}, {Panasenco}, {Titov},
  {Miki{\'c}}, {Reeves}, {Velli}, {Linker}, \& {De Toma}}]{Torok11}
{T{\"o}r{\"o}k}, T., {Panasenco}, O., {Titov}, V.~S., {Miki{\'c}}, Z.,
  {Reeves}, K.~K., {Velli}, M., {Linker}, J.~A., \& {De Toma}, G. 2011, \apjl,
  739, L63

\bibitem[{{Valori} {et~al.}(2005){Valori}, {Kliem}, \& {Keppens}}]{Valori05}
{Valori}, G., {Kliem}, B., \& {Keppens}, R. 2005, \aap, 433, 335

\bibitem[{van Ballegooijen(2004)}]{vanBallegooijen04}
van Ballegooijen, A. 2004, The Astrophysical Journal, 612, 519

\bibitem[{{van Ballegooijen} \& {Cranmer}(2008)}]{vanBallegooijen08}
{van Ballegooijen}, A.~A. \& {Cranmer}, S.~R. 2008, \apj, 682, 644

\bibitem[{{Wheatland}(2006)}]{Wheatland06}
{Wheatland}, M.~S. 2006, \solphys, 238, 29

\bibitem[{{Wiegelmann}(2004)}]{Wiegelmann04}
{Wiegelmann}, T. 2004, \solphys, 219, 87

\bibitem[{Wilmot-Smith {et~al.}(2009)Wilmot-Smith, Hornig, \&
  Pontin}]{WilmotSmith09a}
Wilmot-Smith, A.~L., Hornig, G., \& Pontin, D.~I. 2009, The Astrophysical
  Journal, 696, 1339

\bibitem[{Yang {et~al.}(2015)Yang, Guo, \& Ding}]{0004-637X-806-2-171}
Yang, K., Guo, Y., \& Ding, M.~D. 2015, The Astrophysical Journal, 806, 171

\bibitem[{{Zhao} {et~al.}(2016){Zhao}, {Gilchrist}, {Aulanier}, {Schmieder},
  {Pariat}, \& {Li}}]{Zhao16}
{Zhao}, J., {Gilchrist}, S.~A., {Aulanier}, G., {Schmieder}, B., {Pariat}, E.,
  \& {Li}, H. 2016, \apj, 823, 62

\bibitem[{{Zhao} {et~al.}(2014){Zhao}, {Li}, {Pariat}, {Schmieder}, {Guo}, \&
  {Wiegelmann}}]{Zhao14}
{Zhao}, J., {Li}, H., {Pariat}, E., {Schmieder}, B., {Guo}, Y., \&
  {Wiegelmann}, T. 2014, \apj, 787, 88

\end{thebibliography}
\IfFileExists{\jobname.bbl}{}  
{ 
\typeout{} 
\typeout{****************************************************} 
\typeout{****************************************************} 
\typeout{** Please run "bibtex \jobname" to obtain}  
\typeout{**the bibliography and then re-run LaTeX}  
\typeout{** twice to fix the references!} 
\typeout{****************************************************} 
\typeout{****************************************************} 
\typeout{} 
 }

\end{document}